%% file: paper.tex
\documentclass[acmtog, nonacm]{acmart}

\acmSubmissionID{247}

\setcopyright{acmcopyright}
\acmJournal{TOG}
\acmYear{2022}
\acmVolume{41}
\acmNumber{6}
\acmArticle{245}
\acmMonth{12} 
\acmDOI{10.1145/3550454.3555458}

\setcitestyle{square}

\input{tr-commands}

\myexternaldocument{supplemental}

\newcommand{\app}[1]{\textsc{#1}}

\colorlet{colorGeneral}{black}
\colorlet{colorMeta}{black}
\colorlet{colorFinetune}{black}
\colorlet{colorOverfit}{black}

\usepackage[nolist]{acronym}
\begin{acronym}
\input{acronyms}
\end{acronym}

\input{math}

\def\myparagraph#1{\textit{\textsf{#1.}}\ }

\MakeRobust{\Call}

\begin{document}

\title{Metappearance: Meta-Learning for Visual Appearance Reproduction}

\author{Michael Fischer}
\affiliation{%
	\institution{University College London}
	\country{United Kingdom}
}
\email{m.fischer@cs.ucl.ac.uk}

\author{Tobias Ritschel}
\affiliation{%
	\institution{University College London}
	\country{United Kingdom}
}
\email{t.ritschel@ucl.ac.uk}



\begin{teaserfigure}
   \includegraphics[width=\textwidth]{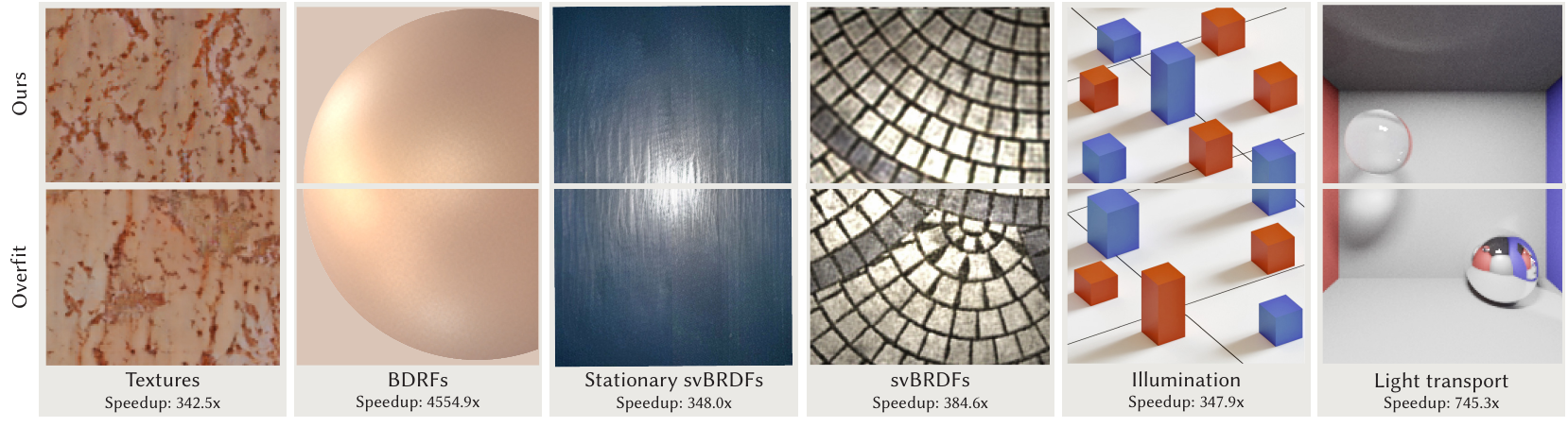}
   \vspace{-.4cm}
   \caption{
   We propose meta-learning for a wide range of appearance reproduction tasks.
   Given as few as 10 optimization steps, our  method (top in each subfigure) achieves quality comparable to overfit-approaches (bottom in each subfigure) that take orders of magnitude more training iterations.
   }
   \label{fig:Teaser}
\end{teaserfigure}

\begin{abstract}
There currently exist two main approaches to reproducing visual appearance using Machine Learning (ML): The first is training models that generalize over different instances of a problem, \eg different images of a dataset. 
As one-shot approaches, these offer fast inference, but often fall short in quality.
The second approach does not train models that generalize across tasks, but rather over-fit a single instance of a problem, \eg a flash image of a material.
These methods offer high quality, but take long to train.
We suggest to combine both techniques end-to-end using meta-learning: We over-fit onto a single problem instance in an inner loop, while also learning how to do so efficiently in an outer-loop across many exemplars.
To this end, we derive the required formalism that allows applying meta-learning to a wide range of visual appearance reproduction problems: textures, \acp{BRDF}, \acp{svBRDF}, illumination or the entire light transport of a scene.
The effects of meta-learning parameters on several different aspects of visual appearance are analyzed in our framework, and specific guidance for different tasks is provided.
Metappearance enables visual quality that is similar to over-fit approaches in only a fraction of their runtime while keeping the adaptivity of general models.
\end{abstract}

\maketitle

\acresetall

\mysection{Introduction}{Introduction}
Reproduction of visual appearance \cite{dorsey2010digital} is a key part of Computer Graphics that has achieved new levels of simplicity, speed and accuracy thanks to recent developments in \ac{ML}.
The classic use of \ac{ML} for appearance reproduction was to capture light or materials from very little input, sometimes only single images \cite{deschaintre2018single,georgoulis2017reflectance}, without access to ground truth maps.
Approaches that are capable thereof usually train for a long time on large datasets and achieve impressive levels of generalization, often due to \acp{CNN} that recognize patterns in the data. 
Unfortunately, this generality comes at the price of not matching the target precisely: we might get a great looking 
\ac{BRDF} or \ac{svBRDF} from a single image, but it might not exactly match the input. 

More recently, a second line of research has evolved, where no attempt is made to generalize over a large dataset, and, instead, non-linear optimization and differentiable rendering are used to explain visual appearance in input images \cite{gatys2015neural,mildenhall2020nerf,rainer2019neural}.
These methods minutely match the reference, but need many input observations, take long to train and can be slow to execute.
Typically, such approaches use point-operations, \eg \acp{MLP}, rather than \acp{CNN}.

A first step to combine these two training paradigms was introduced by adapting the output of a model from the general class in a second, non-end-to-end step, the so-called fine-tuning or post-refinement stage \cite{deschaintre2020guided,henzler2021generative,gao2019deep,guo2020materialgan}.
Approaches that use fine-tuning usually run an additional number of gradient steps (in the order of magnitude $10^3$) towards a specific target, which greatly improves reconstruction quality, but inflates runtime to the order of minutes, whereas feed-forward \acp{CNN} operate in milliseconds.

A dilemma materializes: Should one rather make a user wait in order to provide them with high quality output, or would it be better to provide fast, interactive results that might be of inferior quality? 
Both solutions are unsatisfactory, which is why in this work, we aim to diminish this quality-speed-gap and provide quality that is a) close to model-overfitting or fine-tuning, and b) available at interactive runtimes, close to those of general feed-forward networks. 

\changed{We achieve this by harnessing the power of \emph{meta-learning}: building on the MAML algorithm \cite{finn2017model} from the machine learning community}, our framework Metappearance uses two nested optimization loops, where the outer loop is sequentially presented with all exemplars in a (training) dataset.
For each exemplar, the inner loop is then tasked with over-fitting a model onto this specific exemplar. 
Characteristically, the inner loop operates under the constraint of a very limited number of available gradient descent steps, typically around 10 only. 
Metappearance hence learns to efficiently drive the inner optimization towards a specific target, but still is able to exploit coherency and priors in the data due to knowledge gathered in the outer loop. 

\changed{In this work, we present a framework that formalizes the application of meta-learning to the task of visual appearance reproduction.} 
Importantly, we do not propose new visual appearance methods or new loss functions, nor do we compare methods or analyze their properties. 
In fact, quite the contrary is true: we keep the methods the same, but instead propose a different way of training them. 
\changed{By comparing our approach against ``traditional'' training paradigms, we show which types of applications can benefit from meta-learning and explore the implications on performance and quality.
We validate that Metappearance outperforms mere general inference followed by fine-tuning through ablation- and convergence-studies. 
Additionally, we, for the first time in the graphics literature, make the connection between meta-learning, model compression and data efficiency}.
We show that Metappearance speeds up faithful appearance reproduction by several orders of magnitude, while keeping all desirable properties of the respective base approaches and similar visual quality. 

In summary, our contributions are

\begin{itemize}
    \item Metappearance\footnote{ \changed{Our source code will be available at \url{https://github.com/mfischer-ucl/metappearance.}}}, a model that adapts to new, unseen visual appearance tasks in only a few steps of gradient descent. 
    \item Optimizing for a fast and accurate optimizer of this model. 
    \item Instances of this model that accurately match texture, \acp{BRDF}, \acp{svBRDF}, illumination, or light transport orders of magnitude faster 
    than strong baselines, at comparable quality, and
    \item An analysis of our method's properties, its convergence and its behaviour under ablation.
\end{itemize}

\mysection{Previous Work}{PreviousWork}

\mysubsection{Visual Appearance}{VisualAppearance}
We consider visual appearance reproduction, the task of generating plausible and accurate visual patterns across all positions and orientations from evidence captured for some angles and locations.

Ignoring angle and considering an exemplar's statistics, we would talk about appearance as \emph{texture} \cite{julesz1975experiments,efros1999texture,gatys2015neural}.
When angle matters, we would call this \acf{BRDF} \cite{ngan2005experimental,guarnera2016brdf} and when both space and orientation are considered, \acf{svBRDF} \cite{dana2004device}.
\citet{guarnera2016brdf} summarize these approaches.
Textures \cite{ulyanov2016texture,gatys2015neural}, \acp{BRDF} \cite{georgoulis2017reflectance} and \acp{svBRDF} \cite{deschaintre2018single} have all been acquired and represented by means of \ac{ML}, for which \citet{tewari2020state} provide a survey. 
We defer discussion of the specific existing solutions for all those sub-problems to \refSec{Applications}.

\mysubsection{Learning}{Learning}
More important to our problem is how the different methods are trained, 
\ie optimized, given either the information of a single instance or an entire set of exemplars (\refTab{Methods}).

\begin{table}[htb]
    \setlength{\tabcolsep}{0.3cm}
    \centering
    \caption{Different ways to optimize for visual appearance reproduction.}
    \label{tab:Methods}
    \begin{tabular}{lcccc}
         &
         General&
         Fast&
         Accurate&
         Compact \\
         \toprule
         \method{General}& \cmark & \cmark & \xmark & \xmark \\
         \method{Overfit}& \xmark & \xmark & \cmark & \cmark \\
         \method{Finetune}& \xmark & \xmark & \cmark & \xmark \\
         \method{Meta} (ours)& \cmark & \cmark & \cmark & \cmark \\ 
         \bottomrule
    \end{tabular}
\end{table}

\myparagraph{General Learning}
A typical paradigm is to collect a training dataset, say, 2D images, to curate them with appearance supervision, \eg \ac{BRDF} parameters, and to learn a mapping from the image to those parameters, for example through a \ac{CNN} \cite{georgoulis2017reflectance}.
Often, such methods create a latent space.
While it is a strength that this space will mostly contain valid exemplars, it comes at the expense of a bottleneck, reducing specific details.
In simple words, a 100-dimensional latent space can make sure every latent code is a grass texture, but it cannot represent the exact location of 200 grass blades in an image.
Examples of such approaches include work by
\citet{henzler2020learning} (texture), \citet{georgoulis2017reflectance} (\ac{BRDF}) or
\citet{deschaintre2018single,kuznetsov2019learning,gao2019deep,guo2020materialgan} (svBRDF) or \citet{zhu2020deep,zhu2021photon,bako2019offline,huo2020adaptive} (light paths).
These methods generalize well to new data, but do not \textit{exactly} match the test-time input, and hence are \textit{general} and \textit{fast}, but not \textit{accurate}, as per the taxonomy established in \refTab{Methods}. Moreover, they often require an encoder- and decoder branch, which makes them not \textit{compact}. 
We call these \method{General} and formalize them in \refSec{General}.

\myparagraph{Over-fit Optimization}
A second, more classical approach is to not seek generalization, but to fit a model to samples of a specific problem instance.
This technique has seen a recent increase in popularity due to the emergence of coordinate-based neural representations, and often is used in conjunction with \acp{MLP}. 
Examples are numerous and include most works related to \ac{NeRF} \cite{mildenhall2020nerf} as well as others for texture \cite{kuznetsov2021neumip}, BRDFs \cite{sztrajman2021neural}, svBRDFs \cite{zhang2021nerfactor} or the entire light transport \cite{zheng2019learning,muller2019neural}. 
We call these methods \method{Overfit} and define them in \refSec{Overfit}. Most \method{Overfit} approaches are \textit{accurate} and \textit{compact} (they usually do not require an encoder, as they do not need to generalize), but neither \textit{fast} to train nor \textit{general}. 

\myparagraph{Fine-tuning}
A combination of above approaches is sometimes used, where first a general network is trained and then, when the target instance is known, is optimized a second time \cite{deschaintre2020guided,henzler2021generative}.
Some have employed optimization in latent space \cite{tan1995reducing} while keeping the rest of the network fixed \cite{kang2018efficient,kang2019learning,gao2019deep,guo2020materialgan}, or in pixel space after a user-adjusted number of iterations, aiming to fit the target perfectly.
We here name these \method{Finetune} and define them in detail in \refSec{Finetuning}.
Approaches that use fine-tuning or post-optimization usually are \textit{accurate}, but neither \textit{fast} (post-refinement usually happens at non-interactive runtimes) nor \textit{general} (once the model is fine-tuned, it cannot be used for general inference anymore). Most fine-tuning models are \textit{compact}, as it is usually enough to store the fine-tuned decoder and the corresponding latent code (\cite{henzler2021generative}), although this is not always the case (\cite{deschaintre2020guided}. 

\myparagraph{Hyper- and meta-learning}
Hyper-networks produce weights of another network \cite{ha2016hypernetworks}.
This has been applied to appearance \cite{maximov2019deep,bi2021deep}, \acp{BRDF} \cite{sztrajman2021neural} and \ac{NeRF}-like representations \cite{sitzmann2019scene}.
Meta-learning, instead, does not directly produce the parameters of another network, but guides the optimization that drives the inner learning.
This optimization is often based on gradient descent, so the outer optimization produces setting such as start values and step sizes.
Sometimes, the gradient rule itself is learned \cite{adler2018learned, ravi2016optimization}.
Applications of meta-learning were proposed for 
geometry \cite{sitzmann2020metasdf},
super-resolution \cite{hu2019meta} and animation \cite{wang2021metaavatar},
layered depth images \cite{flynn2019deepview},
as well as for \ac{NeRF} by \citet{bergman2021fast} and \citet{tancik2021learned}.

Approaches that use meta-learning are \textit{fast} and \textit{general} by construction, as they can run inference on new, unseen samples in only a few gradient steps. As we will show in this work, meta-learning for visual appearance reproduction is also \textit{accurate}, as its output is close to overfit- or fine-tuning quality. Moreover, meta-learning enables \textit{compact}ness, as the model initialization and optimization themselves are learned, and hence do not need to rely on latent codes produced by, \eg bulky encoder networks. 
We would not be aware of work attempting to model visual appearance using meta-learning, as we set out to do in \refSec{MetaVersion}.

\mysection{Our Approach}{Method}

After introducing the problem we solve (\refSec{Problem}), we provide a common formalization of three previous solutions (\refSec{General}, \ref{sec:Overfit} and \ref{sec:Finetuning}), and finally introduce Metappearance (\refSec{MetaVersion}).

\mysubsection{Problem statement}{Problem}
We now discuss representing visual appearance, its parametrization, and finally its optimization.

\myparagraph{Representation}
We represent visual appearance as $\radiance_\parameters(\coord|\condition)$, a radiance function of a positional-directional coordinate \coord, conditioned on input \condition and parametrized by the tunable vector \parameters.
The coordinate \coord can be two-, three- or higher-dimensional and might be positional, directional or both.
The condition \condition varies per application and could be a single image, sparse measurements or light paths.

\myparagraph{Parametrization}
Parameterizing $\radiance_\parameters$ by \parameters is possible in a large number of ways, for instance through a plain, pixel-based RGB image, spatial data-structures, or a more implicit representation, like a \ac{CNN} or an \ac{MLP}, and the parametrization might make use of hard-coded, rendering-like operations. 
For now, we deliberately do not specify this further and only require \radiance to be differentiable w.r.t. the parameters \parameters. 
Supplemental Tab. 2 will give examples for instances of this model which we will evaluate in our experiments.

\myparagraph{Optimization}
Let us assume a scalar function $\loss(\parameters,\task)$ that is low if \parameters explains the data \task well and high otherwise.
We will specify different ways to define this loss, leading to different approaches of reproducing visual appearance.
Let us further assume that we have access to an optimizer function  $\learn(\initialParameters,\stepSize,\task,\loss)$ which performs \ac{GD} that starts at  \initialParameters to change parameter \parameters with stepsize \stepSize so as to minimize the loss \loss.
This procedure is given in \refAlg{Learning}, where \stepCount is the number of \ac{GD} iterations.

\begin{algorithm}[htb]
    \caption{Classic learning:
    The function $\texttt{grad}(\cdot)$ differentiates its first argument (an expression) with respect to the second.}
    \begin{algorithmic}[1]
\Procedure{Learn}{\initialParameters, \stepSize, \task, \loss}
        \State \parameters = \initialParameters
        \For{$i\in\{1,\ldots,\stepCount\}$}           
            \State\parameters\ -= \stepSize $\cdot$ \texttt{grad}(\Call{Loss}{\parameters,\task}, \! \parameters) 
        \EndFor
        \State\Return\parameters
    \EndProcedure
    \end{algorithmic}
    \label{alg:Learning}
\end{algorithm}

Combining \loss and \learn leads to different methods.  
In classic learning, the start parameters and the optimization step size both are hyper-parameters that need to be chosen by the user.
We will see that \method{Meta}-learning chooses these optimally through optimization.

\mysubsection{General}{General}
\method{General} methods, that attempt to map a condition \condition directly to appearance, use the loss described in \refAlg{Loss}:

\begin{equation}
\loss_{\method{General}}(\parameters,\task)=
\mathbb E_{i\in\task}[ \changed{ \norm}( \,
\radiance_\parameters
(\coord_i|\condition_i),
\radiance_i \,
)]
\end{equation}

where \changed{$\norm(\cdot\,,\cdot)$} here, and in the following, can refer to any norm.

\begin{algorithm}[htb]
    \caption{Classic loss.
    The function $\sample(\cdot)$ takes a set as an argument and returns a random index into that set.}
    \begin{algorithmic}[1]
\Procedure{Loss}{\parameters, \task}
    \State $cost = 0$
    \For{$i\in\{1,\ldots,\batchSize\}$}
        \State $j$ = \texttt{sample}(\task)
        \State $cost$ += $\norm( \radiance_\parameters(\coord_j|\condition_j), \radiance_j)$
    \EndFor
    \State\Return $cost / \batchSize$
\EndProcedure
    \end{algorithmic}
    \label{alg:Loss}
\end{algorithm}

Visual appearance problems usually are ambiguous: One $\condition_i$ can typically be explained by more than one parameter vector \parameters.
Over the course of training, a \method{General} optimization sees many different conditions $\condition_i$ and hence can build priors about what solutions are more likely than others. 
These priors are then used to generalize to new conditions under new angles and positions, e.g., a new 2D photo of a sphere that can then provide reflectance for new 3D angles and positions.
However, the encoding of these priors that then handle variations over \condition must be performed under the constraint of a finite budget of parameters.
In typical applications, this results in more or less subtle forms of smoothing: a generated \ac{BRDF} does not quite resemble the \ac{BRDF} the input specifies, some spatial details are lost in \acp{svBRDF}, etc.
We will show examples of this in \refSec{Applications}.

\mysubsection{Over-fitting}{Overfit}
Differently, in over-fitting, the loss is
\begin{equation}
\loss_{\method{Overfit}}(\parameters,\task)=
\mathbb E_{i\in\task}[\changed{\norm}( \,
\radiance_\parameters
(\coord_i|\condition),
\radiance_i \,)]
\end{equation}
where, importantly, \condition is constant and does not depend on $i$.
This task is comparatively easy, as the network only has to deal with one specific input. 
Consequently, results are often of higher quality than in the \method{General} setting.
However, the optimization now lacks the synoptic approach that sees all instances and can use this ``bigger picture'' to build priors and make do with fewer information in lower time.
Typically, over-fitting approaches need many iterations to train, take from minutes to hours to converge, and often require further regularization, \eg by physical constraints, to avoid overfitting to specific training positions and directions.

\mysubsection{Fine-tuning}{Finetuning}
Both overfitting and generalization can be combined in a trivial way: First run a general method on the input, second, optimize the output so that it resembles the input even more.
Fine-tuning usually starts from initial parameters \initialParameters, that have been trained across many inputs (\eg the converged state of a general model, cf. \refSec{General}), and then optimizes these for a fixed target, as in over-fitting (\refSec{Overfit}), with a fixed step size \finetuneStepSize. 
This means to compute
\begin{align}
\learn(
\learn(
\initialParameters,
\generalStepSize,
\task,
\loss_{\method{General}}
),
\finetuneStepSize,
\task,
\loss_{\method{Overfit}}
)
.
\end{align}
While the general step can be re-used across several inputs, the subsequent fine-tuning (essentially, over-fitting) optimization must be repeated for each new input. 
\method{Finetune} is faster than \method{Overfit}, as only the inner optimization needs to be executed for inference, while the outer step is a feed-forward network execution, and sometimes is accelerated further by increasing the learning rate $\finetuneStepSize$. 
Still, optimization usually takes in the order of minutes, \ie it is slow compared to a single feed-forward execution of the general network that typically would take milliseconds.
Moreover, the solution might diverge from the priors that informed the first step.
By jointly training over both the general projection and the fine-tuning stage, we overcome these issues in quality and speed, as explained next. 

\mysubsection{Meta-learning}{MetaVersion}
A general model's training is agnostic to the fact that later fine-tuning iterations will be used to further improve the results.
This drives the \method{General} projection step towards learning unnecessarily detailed representations while missing other important features and over-smoothing the space (cf. \refSec{General}).
The \method{General} step hence will try to incorporate features that fine-tuning might include anyway, and subsequently disregard other, more general elements that the fine-tuning operator might miss. 

If we do not know how to trade those properties, could we instead learn how to do that?
Could we learn how to perform an optimization optimally?
To do that, we need i) a domain to optimize over, ii) to understand what is ``optimal'', and iii) an actual algorithm.
We will now look into these aspects.

As our optimization domain, we consider meta learning of both the initial solution \initialParameters as well as a per-parameter step size \stepSize.
Both the initialization and the step sizes are fixed between tasks and stay constant at test time.
We stack \initialParameters and \stepSize into a \emph{meta parameter} vector, denoted as $\metaParameters=\{\initialParameters, \stepSize\}$.
Meta learning can then be formalized\footnote{In a slight abuse of notation, as \learn takes four parameters, while it is called with three here, where the first is a tuple holding the first two arguments, init and stepsize.} as a new loss (\refAlg{MetaLoss}):
\begin{equation}
\loss_{\method{Meta}}(\metaParameters,\metaTask)=
\mathbb E_{i\in\metaTask}[
\loss_{\method{Overfit}}(
\learn(\metaParameters,\task_i,\loss_{\method{Overfit}}), \task_i)
]
.
\end{equation}

The first thing to note is that the loss is defined on meta-parameters \metaParameters and that it calls \learn with these, to quantify how suitable they are for an inner learner. 
Second, it samples from \changed{the space of all tasks \metaTask} (\eg multiple BRDFs), not from a single task.
The sampled task \task is the same for meta-train and meta-test; the same for the call to \learn and to \loss.
Because \sample inside the loss function is randomized, different positions and directions are used for meta-test and meta-train. 
Doing so, parameters that generalize across positions and directions inside one task are advantaged.

To actually perform meta-learning, we
\[\learn(\initialMetaParameters, \metaStepSize, \metaTask, \loss_{\method{Meta}}),\] \ie perform common learning with an advanced loss and a meta-initialization, \initialMetaParameters, as well as a meta step-size \metaStepSize.

\begin{algorithm}[htb]
    \caption{Meta-learning involves a loss that depends on the hyper-parameters of calling the function \textsc{Learn} on the actual task.}
    \begin{algorithmic}[1]
    \Procedure{MetaLoss}{\metaParameters, \metaTask}
        \State $cost$ = 0
        \For{$i\in\{1,\ldots,\batchSize\}$}
            \State $j$ = \sample(\metaTask)
            \State $cost$ += \Call{Loss}{\Call{Learn}{\metaParameters, $\metaTask_j$, \loss}, $\metaTask_j$})
        \EndFor
        \State\Return $cost/b$
    \EndProcedure
    \end{algorithmic}
    \label{alg:MetaLoss}
\end{algorithm}

\myfigure{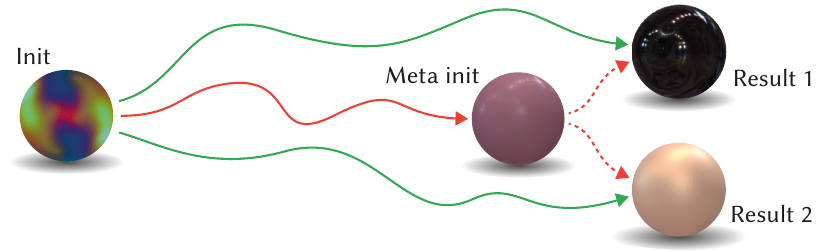}{Learning the init: Trajectories for \method{Meta} and \method{Overfit} for the example task of \app{BRDF} representation.
The dotted line denotes inner optimization.
Note how the dotted trajectories for \method{Meta} are shorter, \ie faster learning.}

By encouraging network parametrizations that enable few-step convergence on unseen samples, meta-learning optimizes over optimization itself.
More specifically, in our scenario, the inner optimizer learns to over-fit to the appearance of one exemplar. 
The outer optimizer then changes the inner optimizer's start parameter values, so that the next inner-loop execution will achieve improved results and do so much quicker.
\refFig{Metalearning} illustrates this idea with two very basic tasks.
As with other losses, the metaloss is computed across a batch, \ie \initialMetaParameters and \metaStepSize are updated with information averaged across multiple optimizations (the \texttt{for} loop in Line 3).

\myfigure{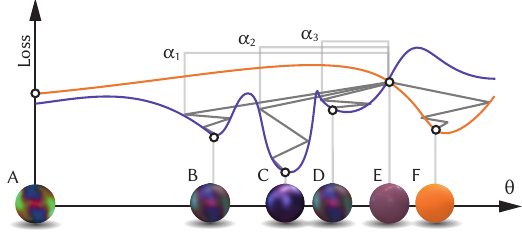}{Learning the step size:
The orange and violet curve show the loss (vertical) for different parameters $\parameters$ (horizontal) for two \ac{BRDF} tasks.
The gray $\alpha\text{-intervals}$ denote three alternative step sizes.
The zig-zags are the convergence paths for specific choices of step size.
Please see the text for discussion.}

\refFig{StepSize} illustrates the purpose of learning the step size.
As explained in \refFig{Metalearning}, meta-learning will change the init from $A$ to a suitable position $E$.
When choosing the step size right, ($\stepSize_2$ for this init) the optimizer will converge to the correct \acp{BRDF}, here $C$ and $F$.
With a step size too small, $\stepSize_3$, or a step size too large, $\stepSize_1$, we converge to less suitable results ($D$ or $B$, respectively) for the violet task.
\changed{As the above considerations might be different in higher dimensions, we parameterize the step size as a vector instead of a scalar, which allows anisotropic gradient steps \cite{li2017meta}.}
During meta-inference, \ie when using the meta-trained model to quickly infer a result for a new, unseen sample provided by a user, the step-sizes are fixed, and only the model weights are changed. 

Jointly learning the model initialization and the corresponding step sizes combines the quality of over-fitting with the ability to build priors of general approaches.
In practice, the inner training loop takes several orders of magnitude fewer iterations than common over-fitting and is up to two orders of magnitudes faster than fine-tuning, which enables execution at interactive rates: in most applications, our inference time is less than 1 second. 

\colorlet{colorGeneral}{colorA}
\colorlet{colorMeta}{colorB}
\colorlet{colorFinetune}{colorC}
\colorlet{colorOverfit}{colorD}
\colorlet{colorRegular}{black}

\paragraph{Implementation}
Our implementation follows the \ac{MAML} framework proposed by \citet{finn2017model}.
As the name suggests, the meta-learner is agnostic to the inner network used, which makes the approach flexible and well-suited for our different application scenarios.
We learn our per-parameter stepsize (cf. \refFig{StepSize}) according to the approach presented in Meta-SGD \cite{li2017meta}. 
For details of the different meta-learning algorithms and tools used, please see Supplemental Sec. 1.

\mysection{Evaluation}{Evaluation}
We have introduced a framework for using meta-learning for visual appearance reproduction, but how well does it compare to more traditional training approaches? 
To answer this, we will now demonstrate the effectiveness of Metappearance on a variety of different applications. 
We will now introduce those, including notes on previous work and the architecture (\refSec{Applications}), then outline the evaluation protocols (\refSec{Methodology}), and, finally, report qualitative and quantitative results (\refSec{Results}).

\mysubsection{Applications}{Applications}
We consider six increasingly complex applications (Supplemental Tab. 2): 
i) RGB textures, 
ii) \acp{BRDF}, 
iii) stationary and 
iv) non-stationary \ac{svBRDF} maps from flash images, 
v) illumination maps from RGB images with normals and finally 
vi) the entire light transport in a scene.

Neither the tasks addressed nor architectures used are novel; the contribution lies in the way they are trained.
We re-iterate that it is hence not our goal to compare different \textit{approaches} (\eg CNN vs. MLP for \ac{BRDF} encoding), but rather compare different methods of \textit{training} a specific approach. 
We will detail each application next.

\mysubsubsection{Textures}{}
In a \app{Texture}, RGB appearance varies over space, but has uniform visual feature statistics \cite{portilla2000parametric}.
\citet{gatys2015neural} optimized for a finite image in pixel space such that its VGG activation statistics match the exemplar, a solution that would be \method{Overfit} in our taxonomy.
Later, \citet{ulyanov2016texture} trained a single \ac{CNN} to perform this task feed-forward.
\citet{huang2017arbitrary} have shown how control over (instance) normalization can produce new textures corresponding to a \method{General} solution in the logic of this work.
\citet{henzler2020learning} show how to do this conditioned on an input image, optionally involving a step of \method{Finetune}.
For a comprehensive survey, we refer the reader to \citet{raad2018survey}.

These methods are exemplary for the spectrum we challenge: either they take long to learn and fit the input exactly, or they are fast and only approximate the input. 
We study a design based on \citet{ulyanov2016texture} and \citet{henzler2020learning} as per \app{Texture} in Supplemental Tab. 2.
For the exact network and training setup, please confer Supplemental Sec. 2.1.

\mysubsubsection{BRDFs}{}
While the RGB textures varied in space, but not in angle, we now look into visual appearance varying with angle, but not over space, the classic \ac{BRDF} representation task.
We use a network the learn the \ac{BRDF} responses for given light- and view-directions. 
Our experiments follow \citet{sztrajman2021neural} and \citet{hu2020deepbrdf}, who both use networks combined with custom parametrizations to encode the MERL \cite{matusik2003data} \ac{BRDF} database.
Details on related work and the architectures used are found in Supplemental Sec. 2.2.

\mysubsubsection{Stationary svBRDFs}{}
The next-higher level of complexity are stationary spatially varying \acp{BRDF} (\app{svBRDFStat}) that combine spatial and angular variation of reflectance, as also surveyed by \citet{guarnera2016brdf}.
The theme recurs: optimization is slow but matches the target well, while feed-forward networks are fast, but often do not reproduce the target.

Specifically, we study estimating stationary \acp{svBRDF} from flash images, pioneered by \citet{aittala2016reflectance}, denoted \app{svBRDFStat} in Supplemental Tab. 2.
We look at a design using a noise-conditioned encoder-decoder, as demonstrated in \citet{henzler2021generative}.
We show re-lit results, parameter maps and all network details and training routines in Supplemental Sec. 2.3.

\mysubsubsection{Non-stationary svBRDFs}{}
Besides stationary \acp{svBRDF}, we look into estimating non-stationary ones (\app{svBRDFNonStat}), also from flash images.
This task was explored by \citet{deschaintre2018single} as well as \citet{guo2020materialgan} and \citet{gao2019deep} before.
They all combine learning with fine-tuning in different ways.
While \citet{deschaintre2020guided} use additional information and upsampling, \citet{gao2019deep} and \citet{guo2020materialgan} optimize first in a latent space, and later in the pixel space given only the target flash image.

We adapt the architecture from \citet{deschaintre2018single}, an encoder-decoder with a re-rendering loss, trained supervised under $L_1$ on synthetic flash images, \app{svBRDFNonStat} from Supplemental Tab. 2.
For testing, both the reference as well as the inferred results are rendered from a set of novel view and light directions and compared. 

\mysubsubsection{Illumination}{}
While the previous applications have looked into different forms of reflectance, another important application for visual appearance is estimating \app{Illumination}.
To study the relation to meta-learning, we consider the task of representing natural spherical illumination itself as a \ac{NN}.
In particular, we consider an encoder-decoder that takes as input a diffuse shaded \ac{LDR} image of a sphere and outputs the \ac{HDR} environment map.
Training data is rendered using the Laval HDR environment map dataset \cite{gardner2017learning} to illuminate spheres of random materials.
For evaluation, we render a second scene under the reference- as well as the inferred illumination and compare both results.
Details are found in Supplemental Sec. 2.5.


\mysubsubsection{Light transport}{SubSecLightTransport}
The ultimate explanation for visual appearance is the light transport in a scene itself \cite{veach1998robust}.
To this day, robust handling of all forms of light transport remains a challenge \cite{keller2020jaroslav}.
One important building block that recently received a lot of attention is the \emph{guidance} of paths \cite{lafortune19955d,vorba2014line,muller2017practical,muller2019neural,herholz2016product,zheng2019learning,zhu2021photon}, where previous paths build a model (parametric or \ac{NN}-based) that is then used to steer path generation towards relevant paths that reduce variance more effectively.
As this strategy involves optimization, it can also be meta-learned, so as to transfering understanding of light transport across scenes.

To do so, we study the architecture of \citet{zheng2019learning}, which relies on a normalizing flow \cite{rezende2015variational} to learn a map from the unit hypercube to \ac{PSS}, such that path density matches the one of the target scene. 
In other words, the normalizing flow learns a scene-dependent \ac{PSS} warp that is then applied to random \ac{PSS} samples. 
Once trained, the model can be used to generate new paths as well as their probability.
For details on architecture and training, please see Supplemental Sec. 2.6.

While previously these methods were applied to a single scene, we consider an entire set of scenes.
To study this, we use a Cornell-box like configuration that is populated by a random number of between 1 and 5 spheres of random diffuse, glass or metal materials in random positions, a mirror that is randomly placed at a sidewall, and an area light positioned randomly on the ceiling.
To quantify success, we measure the \ac{NLL} across the test-set, as is done in \cite{zheng2019learning}, as using image-based metrics to quantify training error would require rendering the entire test-set per training epoch and method, which is computationally intractable. 
If the model matches the scene-dependent radiance distribution well, \ie it correctly adapts to the light transport within the scene, the resulting \ac{NLL} will be low. 
We report image-space metrics with the trained models for an equal number of rays in Tab. \ref{tab:TransportImageMetrics}.

\mysubsection{Methodology}{Methodology}

\myparagraph{Protocol}
We compare our method against the approaches listed in \refSec{Method}.
The protocol is as follows: Let \conditions denote the entire training data set, \eg all \ac{BRDF} samples in the MERL database.
For \method{Overfit}, we sample a single input \condition from \conditions and train a network on \condition, and only \condition, until convergence.
\method{General} denotes a network that has been trained on all elements in \conditions and then is conditioned on the particular \condition we want inference for.
\method{Finetune} applies a pre-defined number of \finetuningSteps fine-tuning steps to the output of \method{General} to further improve the result for a particular \condition.
Finally, our proposed \method{Meta}-learning is trained on all tasks in \conditions, but under the constraint of being given only a fixed budget of \stepCount gradient descent steps, with $\stepCount <\!\!< \finetuningSteps$.
At inference time, a model conditioned on a particular \condition can then quickly be instantiated by updating the initial meta-model parameters \stepCount times. 

\myparagraph{Timing}
All experiments report inference time, \ie the time elapsed between first presenting an input to the network and receiving its final output. 
We refrain from reporting render- or path-tracing times, as these do not change across methods.
Please note that for \method{Meta}, inference time is different from (meta-) training time: meta-inference is quick, as we only need to perform a small number of gradient steps and do not need to calculate costly higher-order derivatives. 
We report this figure, as this is what a user would experience when presenting new, unseen inputs to one of our meta-trained applications. 
During meta-training, however, the opposite is true: we often need to backpropagate though the gradient operator itself, and do so for many examples.
This leads to long meta-training times: \method{Meta} trains roughly twice as long as \method{General}.
For a more detailed listing of training times, cf. Supplemental Tab. 1.
Note that for inference, the speed and memory consumption of the deployed  \ac{NN} is unaffected by meta-learning.

\myparagraph{Metrics}
We evaluate each method on unseen input from the test set, with the particular evaluation metric depending on the application (column ``Metric'' in Supplemental Tab. 2).
In particular, meta-learning does not ``cheat'' by disclosing any test data during training; the split is the same as in conventional training.
This means that \method{Meta} is presented with entirely new tasks (\eg a completely unseen \ac{BRDF}) instead of just withheld samples from a previously processed task (Supplemental Fig. 1 visualizes this). 

\begin{table*}[ht]
\caption{
Quantitative results for all methods on all applications. 
The first row of plots shows the quality-speed continuum spanned by the four methods. 
The ideal range (fast inference \textit{and} high quality) is in the top right corner. 
The second row shows test-set error, individually sorted for each method.
The third row shows convergence plots at inference time. 
Note that very different time-scales are plotted on the same horizontal range, so comparison can only be made in shape, not between fixed values at any point in time.}
    \label{tab:Results}
    \setlength{\tabcolsep}{0.25cm}
\begin{tabular}{l rr rr rr rr rr rr}
    \multicolumn{1}{c}{} &
  \multicolumn{2}{c}{\app{Texture}} &
  \multicolumn{2}{c}{\app{BRDF}}&
  \multicolumn{2}{c}{\app{svBRDFStat}}&
  \multicolumn{2}{c}{\app{svBRDFNonStat}}&
  \multicolumn{2}{c}{\app{Illumination}}&
  \multicolumn{2}{c}{\app{Transport}}
  \\
  \multicolumn{13}c{
  \includegraphics[width=0.97\linewidth]{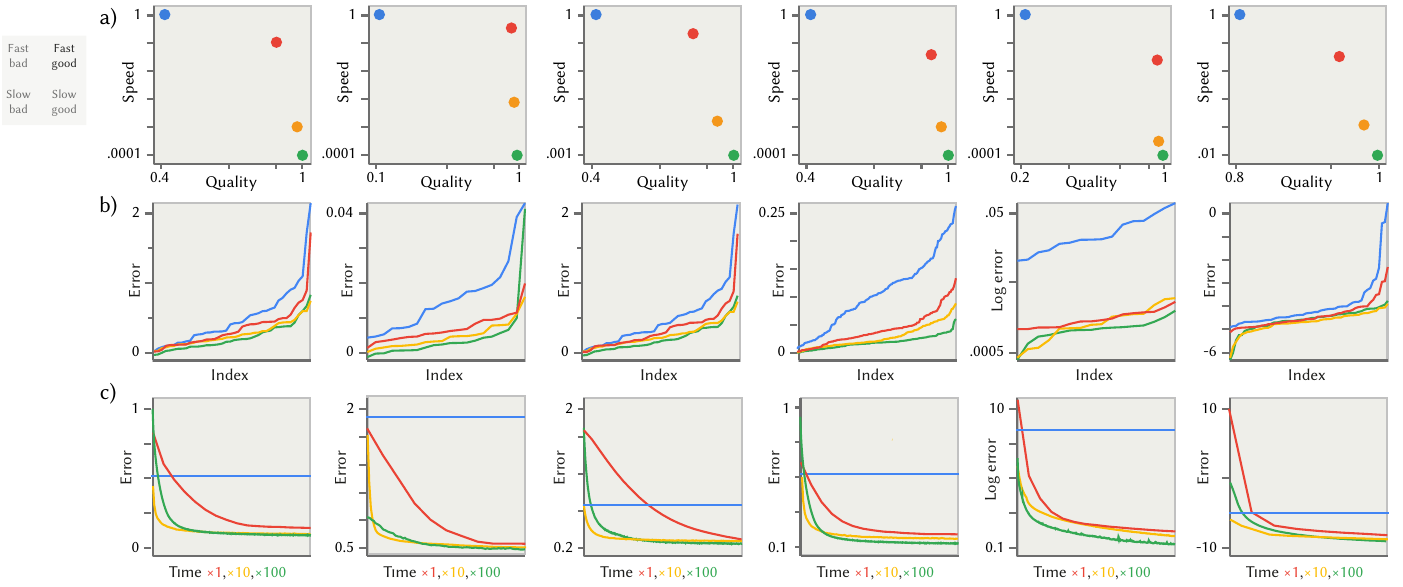}}
\\
\cmidrule(lr){2-3}
  \cmidrule(lr){4-5}
  \cmidrule(lr){6-7}
  \cmidrule(lr){8-9}
  \cmidrule(lr){10-11}
  \cmidrule(lr){12-13}
  &
  \multicolumn1c{Error} &
  \multicolumn1c{Time} &
  \multicolumn1c{Error} &
  \multicolumn1c{Time} &
  \multicolumn1c{Error} &
  \multicolumn1c{Time} &
  \multicolumn1c{Error} &
  \multicolumn1c{Time} &
  \multicolumn1c{Error} &
  \multicolumn1c{Time} &
  \multicolumn1c{Error} &
  \multicolumn1c{Time}
  \\ 
  \midrule
\method{General} & 0.522 & 0.022 & 1.892 & 0.005 & 0.436 & 0.201 & 0.540 & 0.040 & 6.536 & 0.002 & -3.457 & 0.040 \\
\method{Overfit} & 0.183 & 212.421 & 0.631 & 141.201 & 0.229 & 516.515 & 0.099 & 182.292 & 0.168 & 133.583 & -4.120 & 362.210 \\
\method{Finetune} & 0.201 & 21.210 & 0.654 & 9.972 & 0.256 & 103.675 & 0.141 & 57.221 & 0.313 & 26.812 & -4.030 & 42.070 \\
\method{Meta} & 0.252 & 0.619 & 0.720 & 0.031 & 0.311 & 1.484 & 0.197 & 0.474 & 0.352 & 0.384 & -3.970 & 0.486 \\
\bottomrule
\end{tabular}%
\end{table*}

\mysubsection{Results}{Results}
We summarize quantitative results in \refTab{Results}.
We consistently show lower compute time than \method{Finetune} and  \method{Overfit} at only slightly reduced quality.
\changed{The speed-quality plots (\refTab{Results}, a) show that our method is not just a compromise between the speed of \method{General} and the quality of \method{Overfit}, but instead is located much closer to the ideal range (top right corner) than all other methods.} We will now discuss each application's results in turn. 

\mysubsubsection{Textures}{}

\refTab{Results}, b shows distribution of error across the texture task for all exemplars.
We see that while both \method{General} and \method{Meta} struggle more with some specific (not necessarily the same) problem cases, the result is not dominated by outliers.
The progress of training is seen in \refTab{Results}, c, where \method{Meta} achieves a quality more similar to \method{Finetune} than to \method{General}, but in a fraction of the time.

\refFig{ResultsTexture} shows qualitative results that further confirm these quantitative findings.
\method{General} often projects the unseen textures into the latent space only approximately, which results in subtle but noticeable differences in features, color and scale.
\method{Finetune} and \method{Overfit} both almost perfectly replicate the original, as both methods conduct a complete optimization run on the current sample. Our \method{Meta}-method faithfully replicates all textures with only minor differences in style.
For more results, please cf. the supplemental. 

\myfigure{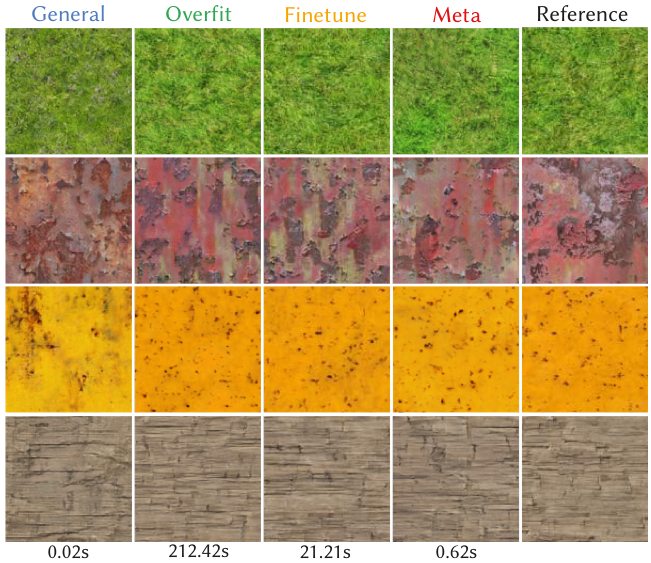}{Results across the test set for the \app{Texture} application. Every result is conditioned on a random process, so not meant to be compared pixel-by-pixel. We report inference time to the respective result.}

\mysubsubsection{BRDFs}{}
Our \method{Meta}-method achieves high fidelity reproduction results across all BRDFs in the MERL-database, as quantified in \refTab{Results}. 
When rendered under Paul Debevec's St.~Peter's Basilica illumination, \method{Meta} achieves \ac{SSIM} values of $\geq 0.95$ on $99\%$ of all materials. 
For a visual comparison of the reproduction results of the different approaches, cf. \refFig{ResultsBRDF}: Our method picks up fine nuances in the \ac{BRDF} correctly, and we consistently show large improvements over \method{General}.
In some cases, \method{Meta} even outperform methods \method{Finetune} and \method{Overfit}, which both have a time budget several orders of magnitudes larger than our method. 

\myfigure{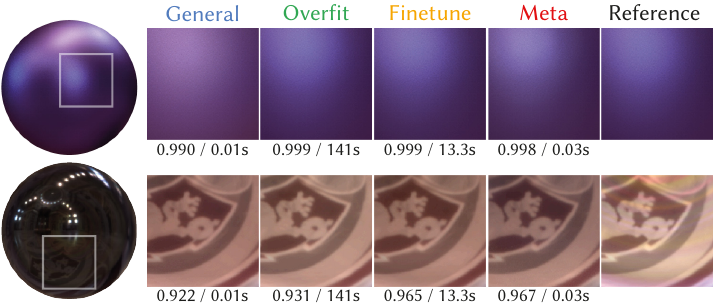}{Rendered results for unseen \acp{BRDF} from the test-set, trained with the different methods. The insets quantify SSIM and inference time.}

\mysubsubsection{Stationary svBRDFs}{}
We detail quantitative results in \refTab{Results} and show qualitative results in \refFig{ResultsSVBRDFstat}. 
This application again confirms our previous findings: \method{General} is fast, but fails to match the input accurately. 
This becomes evident in \refFig{ResultsSVBRDFstat}, where \method{General} broadly matches the target, but is missing fine details and has slightly tinted colors (top and bottom row) or washed-out highlights (middle rows). 
Both \method{Finetune} and \method{Overfit} match the target well and pick up subtle details such as the wood grain and shading cues correctly, but take long to converge.

Our \method{Meta}-method achieves similar visual quality in a fraction of their runtime and manages to produce correct \ac{svBRDF} maps in just over a second. 
We believe that a reason for this is the combination of strong priors built during meta-training (cf. \refFig{MetaIterations}, left column) and learning how to adapt them optimally. 
While the quality-speed improvement is still significant, the gain from using \method{Meta} here is comparatively small, as seen from the position of the red-dot in the quality-speed continuum. 

\myfigure{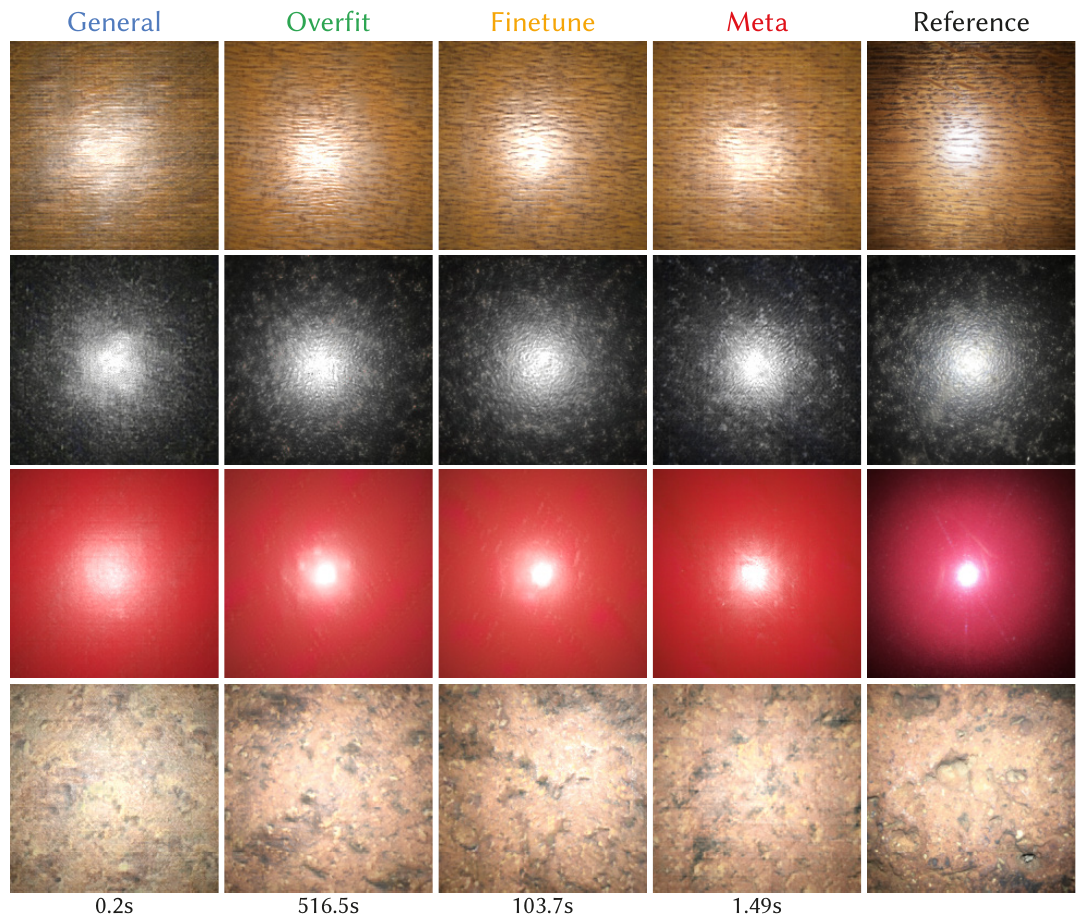}{Results across the \app{svBRDFStat} test-set.
Note that every result is a realization of a random process, so not meant to be compared pixel-by-pixel. For re-lit renderings and shading maps, cf. the supplemental.}

\mysubsubsection{Non-stationary svBDRFs}{}
Qualitative results for non-stationary \acp{svBRDF} are shown in \refFig{ResultsSVBRDFnonstat}.
We see that \method{General} is producing highlights not present in optimization-based training schemes, including ours. 
Out of those optimization-based methods, ours is several orders of magnitude faster, as seen in the numbers and the speed-quality plot in \refTab{Results}, column ``svBRDFNonStat''.
Both \method{Finetune} and \method{Overfit} perform well, although the visual comparison in \refFig{ResultsSVBRDFnonstat} shows \method{Finetune} to perform slightly better. 
We presume that this is due to the fact that \method{Finetune} benefits from the priors developed by \method{General} (recall, fine-tuning starts at the output of the general model), whereas \method{Overfit} starts the training from scratch.
In heavily ill-posed tasks like \ac{svBRDF} estimation, the importance of solid priors has been shown to be of great importance for the optimization (cf. \cite{guo2020materialgan, gao2019deep}). 
This is also part of the explanation for \method{Meta}'s success on this task, as it can build priors over the dataset in the outer loop \textit{and} perform a quick overfit-optimization in the inner loop without diverging too far. 

\myfigure{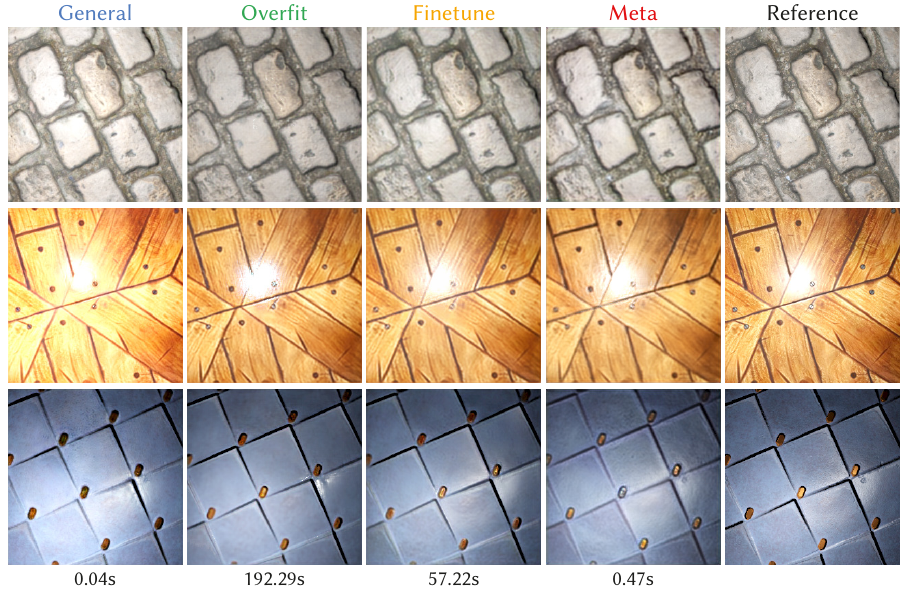}{Relighting results across the test-set for the \app{svBRDFNonStat} task. We render the resulting parameter maps under a different view- and light angle. For more results and the parameter maps, cf. the supplemental.}

\mysubsubsection{Illumination}{}
We present results for our \app{Illumination} task in \refFig{ResultsIllumination}, where we render the inferred envmaps on a scene with a specular and diffuse object. 
We compare all methods against a reference image of that same scene rendered under the groundtruth illumination.
We note how \method{General} is able to place sharp shadows (indicating it handles HDR well) but does not manage to exactly match the intensity.
Similarly, reflections look plausible, but do not match the reference.
Optimization-based methods meet this requirement, but only our meta-trained approach is orders of magnitude faster and achieves comparable quality. 
This is confirmed by the quality-speed plots in \refTab{Results}, where \method{Meta} is far-right, indicating that quality is very close to the full optimization-based methods.
\refTab{Results}, c shows, that \method{Meta} converges even faster than for other tasks (the red curve is more concave).
From the distribution in \refTab{Results}, b we see that the classic optimization-based methods have no problematic outliers, a property retained by \method{Meta}.

\mycfigure{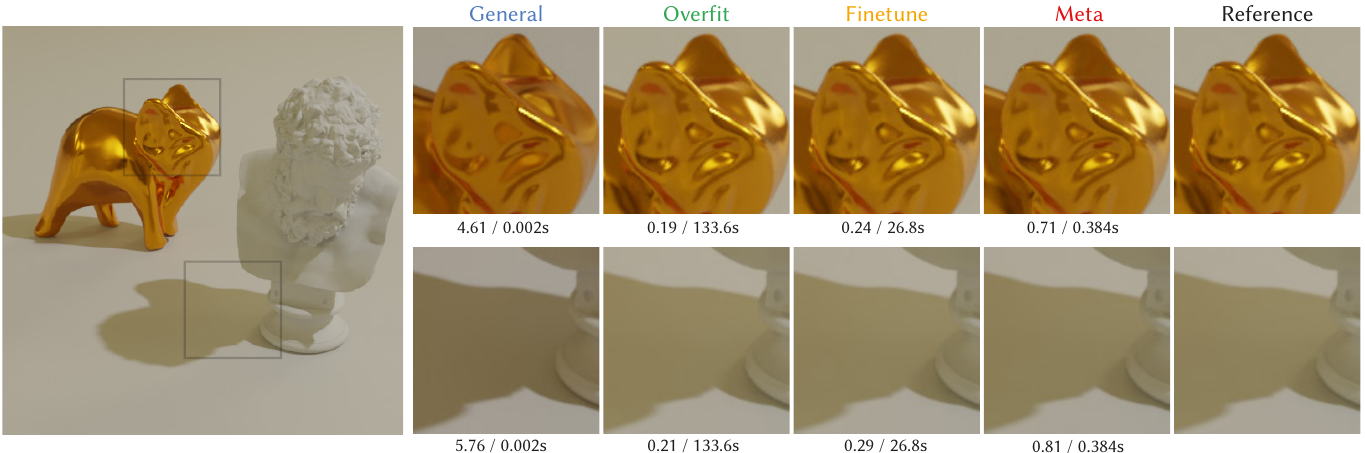}{Results for an unseen instance from the \app{Illumination} test-set, inferred from a single RGB image (not shown) and used to render a novel scene (left).
Quality of illumination is most revealed in reflections (top) and cast shadow (bottom).
The sharpness and shape of the shadows is very indicative of the high dynamic range of the regressed envmap.
We report MAE $\times 10^2$ and inference time.
For direct visualization of the envmaps, cf. the supplemental.}

\mysubsubsection{Light transport}{}
We show the outcome for \app{Transport} in \refFig{ResultsTransport}.
Recall that \app{Transport} uses a resampling of a scene's radiance distribution to learn a model that is used for importance sampling that particular scene.
To compare the effectiveness of each approach, we render a novel scene (unobserved during training for \method{General} and during meta-training for \method{Meta}) using samples generated by the respective importance-sampling model. 
We further include an additional baseline, \method{Regular}, for this application. \method{Regular} uses importance-sampling for the geometric term and randomly samples outgoing paths from the hemisphere oriented around the normal. 
All subfigures are rendered with the same number of samples (4096) and very similar compute time, as querying the importance model can be parallelized on the GPU and hence is fast compared to the tracing of rays (milliseconds vs. minutes).
As elaborated earlier (cf. \refSec{SubSecLightTransport}), we report the time between model instantiation and the final training step (\ie when it is ready to produce light path samples) as this is the part that the choice of training scheme can influence, whereas the subsequent path-tracing time is  approximately invariant\footnote{We write approximately as the \ac{PSS} samples created by all \textit{trained} methods are created in Python and must be passed to the C++ renderer, which incurs a time overhead that the \method{Regular} baseline does not suffer. However, this is in the order of milliseconds, and hence can be neglected.} to the origin of the samples. For a comparison of ray-generation and -tracing times, cf. Supplemental Tab. 4. 

Quantitative results are seen in \refTab{Results}, column \app{Transport}: \method{Meta} is again closest to the top-right corner, indicating that it can combine the quality of \method{Overfit} and the speed of \method{General}.
The qualitative outcome (\refFig{ResultsTransport}) indicates that all methods successfully reduce variance w.r.t. the \method{Regular} baseline.
Again, \method{General} performs slightly below the iterative optimization-based methods, and again, \method{Meta} is orders of magnitude faster. 
To quantify how well the instantiated models perform in image space, we rendered the entire test-set with light paths produced by the respective importance model and a fixed sampling budget of 1024 rays per pixel. We display common metrics calculated on these renderings in Tab. \ref{tab:TransportImageMetrics}.
The results confirm the qualitative inspection in \refFig{ResultsTransport} and show that \method{Meta} again is much closer to \method{Overfit} and \method{Finetune} than to the general or regular baseline. For details on the sampling and rendering operations, we again refer to Supplemental Sec. 2.6.

Please note that we explicitly refrain from discussing which method of importance sampling or path guiding is most appropriate for practical applications and re-iterate that we compare ways of \textit{training} approaches instead of directly comparing the performance of different approaches. We here introduce meta-learning to the importance-sampling and rendering community as a first proof of concept and show that a meta-importance-sampler can generalize across a distribution of Cornell box-like scenes with practical benefits. To our knowledge, this is the first application of meta-learning in rendering, and the first presentation of meta-learned normalizing flows. 

\begin{table}[]
\setlength{\tabcolsep}{0.24cm}
\caption{Mean absolute percentage error and \ac{DSSIM} across the \app{Transport} test-set. Lower is better for both metrics.}
\begin{tabular}{@{}lrrrrr@{}}
         & \method{Regular} & \method{General} & \method{Overfit} & \method{Finetune} & \method{Meta}  \\ \midrule
MAPE     & 0.516   & 0.471   & 0.405   & 0.409    & 0.422 \\
DSSIM & 0.546   & 0.479   & 0.418   & 0.425    & 0.430 \\ \bottomrule
\end{tabular}
\label{tab:TransportImageMetrics}
\end{table}

\mycfigure{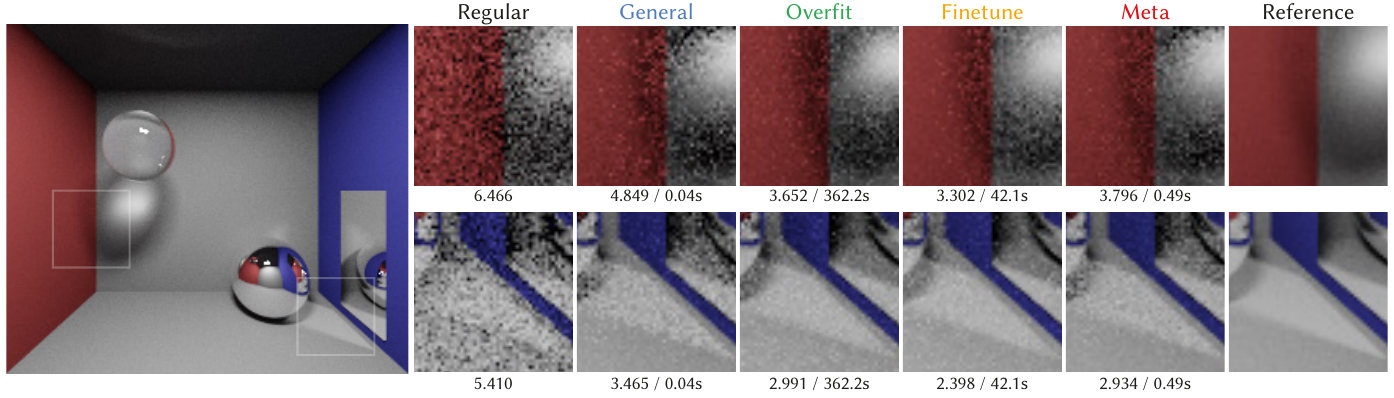}{Results for an unseen test-scene for the \app{Transport} application. 
The left image is rendered with 240,000spp to guarantee a noise-free reference.
The images to the right are produced by rendering the reference scene with \ac{PSS} samples produced by our different approaches (equal number of samples, \ie equal rendering time). We report symmetric mean absolute percentage error (SMAPE) and the respective model inference time.}

\mysection{Analysis}{Analysis}

As the previous section has shown, \method{Meta} achieves similar quality than optimization-based approaches that take orders of magnitude more training time. To analyze the inner workings of Metappearance, we will next discuss a range of further properties that can be deduced from our experiments:
We will ablate our learned components (\refSec{Ablations}), look at convergence of the inner and outer loop (\refSec{Convergence}), explain how Metappearance can be interpreted as model compression (\refSec{Storage}) and finally discuss how \method{Meta} can make do with much fewer input observations (\refSec{SampleEfficiency}).

\mysubsection{Ablations}{Ablations}
\myfigure{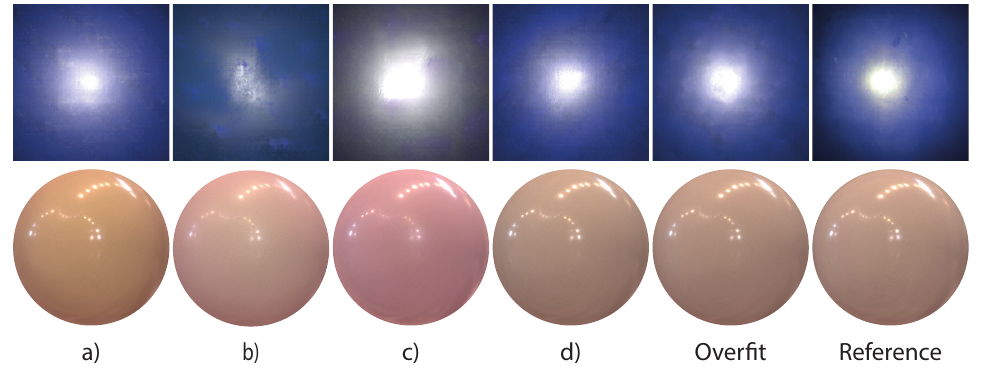}{We compare the influence of a learned learning rate (column b) and initialization (column c). Cf. the main text for details.}
We meta-learn two hyper-parameters: step size and inititalization, but which of them actually contributes to the success?
\refFig{Comparison-FT-LR} looks into this question.
In column a), we display the output of \method{General}. 
In column b), we take this as starting point and use our learned step size to perform $\stepCount=20$ ``smart'' gradient steps towards the reference. 
Evidently, this leads to inferior results, as the general model has been trained with a fixed, global learning rate, and hence does not know how to account for a per-parameter learning rate.
In column c), we use our meta-learned initialization to perform $\stepCount=20$ steps of conventional Adam optimization (learning rate multiplied by 10 for faster convergence) towards the reference. 
This again leads to poor results, as our learned initialization normally is adapted through large, non-uniform gradient steps.
During meta-training, the initialization was hence moved to a region of the objective space that is approximately equally well-suited for all tasks, but not necessarily easy to navigate with uniform gradient steps.
Column d) finally shows the output of our \method{Meta}-method, where learned initialization and step size are used in combination. 
Evidently, this outperforms all alternative configurations. 

This decay in reproduction quality shows that meta-learning really combines the best of both worlds: By optimizing for optimization directly, the outer optimizer can discover gradient paths that lead to local minima by not only moving the network weights (as would \method{General}), but also the step-size with which these weights are updated for a certain number of iterations (as in \method{Finetune}). 

It is tempting to argument that optimizing over optimization itself, \eg learning the step size, automates time-consuming hyperparameter searches. While this is true to a certain extent, one still must choose the \textit{meta}-hyperparameters, \eg outer loop learning rate, etc.  
All our experiments use very similar hyper-parameters (cf. Supplemental Tab. 1) that were not particularly tuned, but this might be different in different applications or designs (cf. \cite{antoniou2018train}). 

\mysubsection{Convergence}{Convergence}
\myfigure{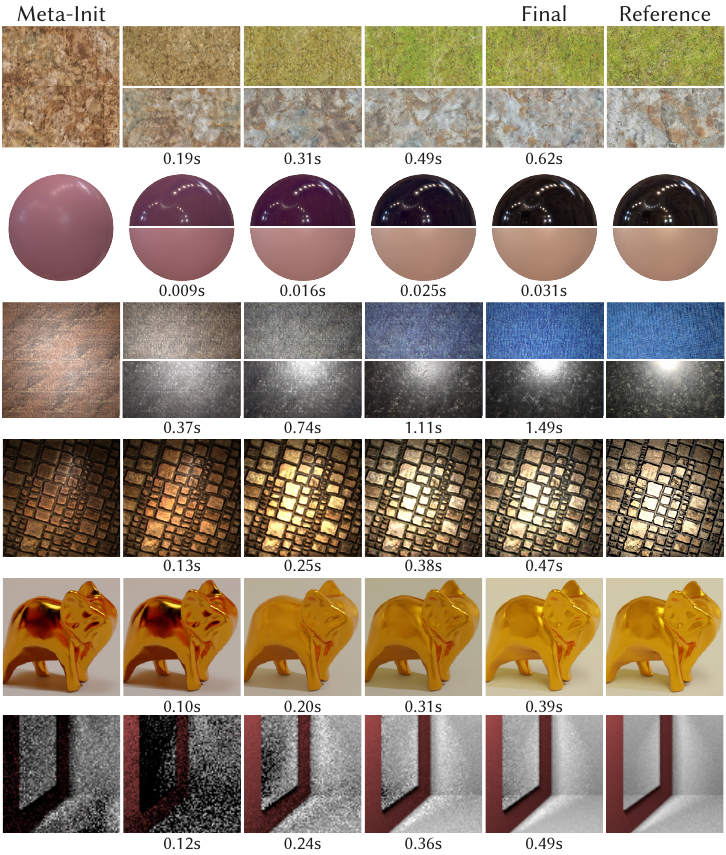}{
Convergence on unseen test-tasks from the meta-init (left) to different targets (right) for our different applications.
We report wall-clock inference time and results after approx. 25\%, 50\% and 75\% (columns 2 - 5) of the inner-loop steps. 
Note that the init itself is a plausible instance and enables the optimization to ``branch'' to specific, very different goals. 
The result for \app{Transport} is an equal-spp rendering with the inferred importance model. Note how the noise clears with more meta-iterations although the sample count stays the same.}

In \refFig{MetaIterations}, we show the convergence of our meta-learned initialization (leftmost column) towards different targets. 
All intermediate outputs show realistic appearance, and even the meta-initialization could pass as a problem instance (\eg a texture) on its own. 
Throughout optimization, our method does not introduce unwanted artifacts, even for the  ambiguous single-image \ac{svBRDF} estimation task.

For maximal quality, we can fine-tune a converged \method{Meta} model for an additional number of training steps. We explicitly refrained from doing so in our main experiments, as this defies the purpose of Metappearance (achieving high quality \textit{without} fine-tuning). However, \method{Meta} will converge to a loss difference of less than 1\,\%  relative to \method{Overfit} in only 65 additional training steps (less than 5 seconds on every application).
Related, \method{General} will not improve from further general training and has been trained to saturation already.

\changed{However, it is not our objective to claim superior quality for infinite time and compute resources.
Instead, let us consider what often is the case in applications that involve user interaction: operating under the constraint of a finite time budget. Imagine, for instance, an architect wanting to quickly add a real-world svBRDF to his 3D model; imagine an app instantaneously adding a customized deep visual appearance model to a Tik-Tok video.
The results of these applications must not only be highly accurate, but also available within split seconds to keep the user engaged, and no other method comes close to ours in this quality-speed trade-off. 
We hence claim highest quality under the constraint of limited time, and}
additionally investigate an equal-time comparison between \method{Finetune} and \method{Meta}, where both approaches are allowed to perform the same number of gradient steps \stepCount that would normally be used during meta-inference. This effectively is a very quick fine-tuning session, which is why we refer to it as ``QuickFinetune'', or \textbf{QuickFT}\footnote{To be able to fine-tune, we need to run \method{General} once. We do not deduct this runtime from \textbf{QuickFT}'s time budget, as this makes the comparison stronger and also usually is a rather fast operation.}. \textbf{QuickFT} uses Adam, for which we again increase the learning rate for faster convergence, as we have done for almost all applications in our experiments (the \textbf{QuickFT} config used here is the same as in Supplemental Tab. 1), while \method{Meta} uses both its learned init and the learned stepsize. 

As \refTab{QuickFinetune} shows, \textbf{QuickFT} already offers great gains over the general method on some applications. However, it is outperformed by \method{Meta} on all applications. \changed{For example-specific visualizations of this experiment, including equal-time comparisons to \method{Overfit}, cf. Supplemental Fig. 6}. This confirms that meta-learning is more than mere general inference followed by fine-tuning and that optimizing over the optimization procedure itself (recall, we learn how to overfit a sample efficiently) really finds a non-trivial optimizer. 

\begin{table}[]
\setlength{\tabcolsep}{0.35cm}
\caption{Average error across the respective application's test-set for our QuickFinetune-experiment. For the metrics reported, please cf. Supplemental Tab. 2. For convenience, we repeat results for methods \method{General} and \method{Meta} from \refTab{Results}.}
\begin{tabular}{@{}lrrrr@{}}
  & \stepCount & \method{General} & \textbf{QuickFT} & \method{Meta}  \\ \midrule
\app{Texture}       & 15 & 0.522   & 0.285 & 0.252              \\
\app{BRDF}          & 10 & 1.892   & 1.346 & 0.720              \\
\app{svBRDFStat}    & 20 & 0.436   & 0.351 & 0.311              \\
\app{svBRDFNonStat} & 15 & 0.540   & 0.289 & 0.197              \\
\app{Illumination}  & 15 & 6.536   & 5.240 & 0.352              \\
\app{Transport}     & 8 & -3.457   & -3.551 & -3.970              \\ \bottomrule
\end{tabular}
\label{tab:QuickFinetune}
\end{table}

\mysubsection{Compression and Efficiency}{Compression}
\mysubsubsection{Storage}{Storage} Our \method{Meta}-method can further be interpreted as a compression scheme.
Consider a rendering or 3D-modeling application, \eg Blender, that loads a pre-trained model's weights to create new, diverse textures for surfaces. 
With methods \method{Overfit} and \method{Finetune}, such an application would have to store an entire set of weights per texture to be generated (note that, in such a scenario, storing the fine-tuned decoder of method \method{Finetune} is sufficient), and then is restricted to synthesizing the pre-learned texture exemplars. 
One could alternatively store the heavy-weight general model, but then would either have to forego accurate high-quality synthesis or fine-tune the result, which is unsatisfying and time-consuming, respectively. 
With our proposed \method{Meta}-method, it is sufficient to store two sets of weights only (the model's weights, and the per-parameter learning rate) to achieve high-quality, diverse texture synthesis in interactive runtime, \ie in less than a second. 

Similar arguments can be made for all the applications we have presented in this work. Let \modelsize denote the number of disk space required to store the model's weights in methods \method{Overfit} and \method{Finetune} (we omit \method{General} from this comparison as we are concerned with high-quality results only). 
The total amount of storage required for the efficient and exact synthesis of \trainingSetSize textures then equals $\modelsize\trainingSetSize$, \ie one set of weights per exemplar. 
Our \method{Meta}-method achieves similar-quality results with \textit{constant} storage requirement $2\modelsize$ (weights and per-parameter learning rate), and is not limited to pre-trained or fine-tuned texture exemplars but instead can quickly infer new, unseen exemplars with high fidelity. 
The compression factor our method achieves hence is $2\trainingSetSize^{-1}$, which, in the case of our exemplar texture application with 500 exemplars, equals $1:250$. 

We would furthermore like to point out that compression also is an inherent property of using neural networks on certain problem instances. 
An NBRDF (a \ac{BRDF} encoded in a network, cf. \cite{sztrajman2021neural}), for instance, has a significantly smaller memory footprint than regular \acp{BRDF} (28\,kB vs. 34.2\,MB in the case of our \method{Meta}-NBRDF), so we additionally compress the original \ac{BRDF} with a factor of approx. $1 : 2000$. 
However, we do not claim credit for this as a property of our method, but rather of the base approaches we meta-train. In fact, quite the contrary is true -- our method requires double the storage of a single network (model weights and per-parameter learning rate). However, we believe \method{Meta}'s ability to quickly converge to unseen tasks and the resulting, aforementioned compression arguments to outweigh this moderate increase. 

\mysubsubsection{Sample Efficiency}{SampleEfficiency}
For methods that consume single data points, \ie methods that use an \ac{MLP}, there is a further compression argument that can be made. To do so, we would like to draw attention to the way \method{Meta} is trained. 

Recall that in meta-learning, the inner loop performs a fixed (small) number of gradient descent steps towards the reference. Naturally, as with most recent optimization algorithms, \method{Meta} uses \textit{stochastic} gradient descent, \ie the inner-loop gradients are not calculated per-sample or across all samples, but rather over a randomly selected subset of all available samples. Evidently, as \method{Meta} only performs \stepCount inner loop steps, only \stepCount batches of size $b$ are sampled. 

Naturally, this process repeats many times during meta-training and hence eventually samples all data in a task (\eg all samples in a \ac{BRDF}). However, this is not the case during meta-inference: Recall that for meta-inference, we merely run \stepCount gradient steps on a new, unseen task. Evidently, this leads to \method{Meta} seeing only $\stepCount \times b$ samples of a task, while all its competitors have access to \textit{all} the samples in a task -- in the cases of \method{Overfit} and \method{Finetune}, even repeatedly. Nonetheless, \method{Meta} delivers quality that is very close to its competitors that have seen all data. 

This is no surprise: The ability to make do with scarce data is a core property of meta-learning and has been amply explored in previous works \cite{al2021data, finn2017model}.
In Metappearance, this property could be useful in a number of ways: Imagine, for instance, a client-server architecture, where the server stores large amounts of data (\eg a large set of scene-dependent \app{Transport} radiance distributions), and the client stores the neural representation that will be trained on samples of this data.
In order to train a model on a specific radiance distribution, the server would have to transfer \textit{all} of its samples to the client (let us ignore the considerable time cost of training or fine-tuning a network on this data for a second). However, following the above argument, we only need to transfer $\stepCount \times b$ samples to infer a converged instance of \method{Meta}, for which the inference time can \textit{really} be neglected.

In summary, this higher efficiency leads to bandwidth savings of approx. 72.2\% for the \app{Transport} application (a full dataset is $288,\!000$ samples, \method{Meta} consumes only $8\times10,\!000=80,\!000$ samples). For the case of \app{BRDF} encoding, the resulting reduction in needed data transmission is even more drastic: \method{Meta} takes \stepCount=10 inner loop steps with a batchsize of $b=512$ and hence consumes 5,120 samples, whereas a full MERL \ac{BRDF}, as is needed by all other methods, consists of $180\times90\times90 \approx 1.46\times10^6$ samples. \method{Meta} hence achieves a bandwidth saving of $99.6$\%.

\mysection{Conclusion}{Conclusion}
We have used meta-learning for efficient and accurate appearance-reproduction on a variety of increasingly complex applications. 
Our model, Metappearance, provides users with results that qualitatively compare well to other training schemes which take orders of magnitude more training iterations or data.
We have shown that Metappearance generalizes not only across problem instances of a similar nature, \eg our variety of Cornell-box scenes, but can also be applied across applications.
In terms of implementation effort, the additional code relative to a solution that already uses an existing optimization is small.
In fact, as we have shown in \refSec{MetaVersion}, re-phrasing the loss function is sufficient.

The main point of our experimentation is that while we cannot yet have both perfect speed and perfect quality, we, in several cases, improve substantially over a mere trade-off between the two, as seen from the red dot in \refTab{Results}, a), which has moved much closer to the ideal top-right spot, where visual appearance reproduction aims to be. 
Directions for future research could include the application of Metappearance to even more complex light-transport algorithms or its extension to meta-learning the objective function or the sampling pattern itself, which would enable even higher accuracy for visual appearance reproduction.

\begin{acks}
This work was supported by \grantsponsor{MRL}{Meta Reality Labs}{}, Grant Nr.\ \grantnum{MRL}{5034015}.
We also acknowledge Gilles Rainer, Alejandro Sztrajman and Philipp Henzler for proofreading.
\end{acks}


\bibliographystyle{ACM-Reference-Format}
\bibliography{paper}

\end{document}

%% file: tr-commands.tex
\usepackage{soul}
\soulregister\ref7
\soulregister\cite7
\soulregister\citeetal7
\soulregister\refSec7
\soulregister\refFig7
\soulregister\refTbl7
\soulregister\refEq7
\soulregister\cite7
\soulregister\ref7
\soulregister\pageref7
\soulregister\shortcite7
\soulregister\eg7
\soulregister\ie7
\soulregister\etal7
\soulregister\unsure7

\usepackage{microtype}
\usepackage{multicol}
\usepackage{amsmath}
\usepackage{amsfonts}
\usepackage{booktabs}
\usepackage{hyperref}
\usepackage{multirow}
\usepackage{graphicx}
\usepackage{dsfont}
\usepackage{pifont}
\usepackage{xcolor}
\usepackage{wrapfig}
\usepackage{nicefrac}
\usepackage{flushend}
\usepackage{ctable}
\usepackage{natbib}
\usepackage{xspace}
\usepackage{algorithm}
\usepackage{algpseudocode}
\usepackage{comment}
\usepackage{soul}

\usepackage[normalem]{ulem}
\usepackage{caption}
\usepackage{subcaption}

\DeclareGraphicsExtensions{.png,.jpg,.pdf,.ai,.psd}
\graphicspath{{images/}}

\def\myfigure#1#2{%
    \begin{figure}[tb]%
    \centering\includegraphics*[width = \linewidth]{#1}%
    \vspace{-.3cm}%
    \caption{#2}%
    \label{fig:#1}%
    \end{figure}%
}

\def\mycfigure#1#2{%
    \begin{figure*}[htb]%
    \centering\includegraphics*[width = \linewidth]{#1}%
    \vspace{-.2cm}%
    \caption{#2}%
    \label{fig:#1}%
    \end{figure*}%
}

\newcommand{\mysection}[2]{\section{#1}\label{sec:#2}}
\newcommand{\mysubsection}[2]{\subsection{#1}\label{sec:#2}}
\newcommand{\mysubsubsection}[2]{\subsubsection{#1}\label{sec:#2}}

\newcommand{\refSec}[1]{Sec.~\ref{sec:#1}}
\newcommand{\refFig}[1]{Fig.~\ref{fig:#1}}
\newcommand{\refEq}[1]{Eq.~\ref{eq:#1}}
\newcommand{\refTab}[1]{Tab.~\ref{tab:#1}}
\newcommand{\refAlg}[1]{Alg.~\ref{alg:#1}}

\newcommand{\ie}{i.e.,\ }
\newcommand{\eg}{e.g.,\ }

\newcommand{\mymath}[2]{\newcommand{#1}{\TextOrMath{$#2$\xspace}{#2}}}

\newcommand{\unsure}[1]{{\sethlcolor{yellow}\hl{#1}}}

\definecolor{colorA}{HTML}{4285f4}
\definecolor{colorB}{HTML}{ea4335}
\definecolor{colorC}{HTML}{ff9900}
\definecolor{colorD}{HTML}{34a853}
\definecolor{colorE}{HTML}{ff6d01}
\definecolor{colorF}{HTML}{46bdc6}
\definecolor{colorG}{HTML}{000000}
\definecolor{colorH}{HTML}{777777}
\definecolor{neonYellow}{HTML}{ffff00}
\definecolor{neonGreen}{HTML}{abff00}

\newcommand{\method}[1]{\textcolor{color#1}{\texttt{\textbf{#1}}}}

\sethlcolor{neonGreen}

\newcommand{\changed}[1]{\textcolor{black}{#1}}

\usepackage{siunitx}
\usepackage{caption}

\usepackage{pifont}
\newcommand{\cmark}{\checkmark}%
\newcommand{\xmark}{\scalebox{0.85}{\ding{53}}}

\usepackage{xr}
\makeatletter
\newcommand*{\addFileDependency}[1]{
  \typeout{(#1)}
  \@addtofilelist{#1}
  \IfFileExists{#1}{}{\typeout{No file #1.}}
}
\makeatother

\newcommand*{\myexternaldocument}[1]{%
    \externaldocument[supplemental-]{#1}%
    \addFileDependency{#1.tex}%
    \addFileDependency{#1.aux}%
}

%% file: acronyms.tex
\acro{BRDF}{Bi-directional Reflectance Distribution Function}
\acro{BTF}{Bidirectional texture function}
\acro{BN}{Batch normalization}
\acro{CNN}{Convolutional neural network}
\acro{FOMAML}{First-Order MAML}
\acro{HDR}{High-dynamic range}
\acro{LDR}{Low-dynamic range}
\acro{GD}{Gradient descent}
\acro{GMM}{Gaussian mixture model}
\acro{MAE}{Mean absolute error}
\acro{MC}{Monte Carlo}
\acro{ML}{Machine learning}
\acro{MLP}{Multi-layered perceptron}
\acro{MAML}{Model-agnostic meta-learning}
\acro{NeRF}{Neural radiance fields}
\acro{NN}{Neural network}
\acro{PCA}{Principal component analysis}
\acro{svBRDF}{spatially-varying BRDF}
\acro{SSIM}{Structural Similarity}
\acro{PSS}{Primary Sample Space}
\acro{PDF}{Probability Density Function}
\acro{DSSIM}{Structural Dissimilarity}
\acro{NLL}{Negative Log-Likelihood}

%% file: math.tex
\mymath{\expecation}{\mathbb E}
\mymath{\conditions}{\mathcal I}
\mymath{\condition}{I}
\mymath{\parameters}{\theta}
\mymath{\initialParameters}{{\parameters_0}}
\mymath{\radiance}{L}
\mymath{\coord}{\mathbf x}
\mymath{\batchSize}{b}
\mymath{\task}{T}
\mymath{\metaTask}{{\mathcal T}}
\mymath{\argmin}{\operatorname{arg\,min}}
\mymath{\trainingSetSize}{m}
\mymath{\stepSize}{\alpha}
\mymath{\generalStepSize}{{\stepSize_\mathrm{G}}}
\mymath{\overfitStepSize}{{\stepSize_\mathrm{O}}}
\mymath{\finetuneStepSize}{{\stepSize_\mathrm{F}}}
\mymath{\metaStepSize}{{\stepSize_\mathrm{M}}}
\mymath{\metaParameters}{\phi}
\mymath{\initialMetaParameters}{\phi_0}
\mymath{\metaBatchSize}{c}
\mymath{\stepCount}{n_\mathrm l}
\mymath{\metaStepCount}{n_\mathrm m}
\mymath{\modelsize}{w}
\mymath{\finetuningSteps}{n}
\mymath{\learn}{\textsc{Learn}}
\mymath{\loss}{\textsc{Loss}}
\mymath{\sample}{\texttt{sample}}
\mymath{\norm}{\Delta}

\mymath{\innerloopsteps}{k}